\documentclass{osa-article}

\journal{osajournal}

\usepackage{graphicx}
\usepackage{comment}
\articletype{Research Article}

\begin{document}

\title{Numerical Analysis of a Highly Sensitive Surface Plasmon Resonance Sensor for SARS-CoV-2 Detection}

\author{Syed Mohammad Ashab Uddin,\authormark{1,4,+}Sayeed Shafayet Chowdhury,\authormark{2,5,+} and Ehsan Kabir\authormark{3,6}}

\address{\authormark{1}Department of Electrical Engineering and Computer Sciences, University of California,  Irvine, CA 92697, USA\\
\authormark{2}Department of Electrical and Computer Engineering, Purdue University, West Lafayette, IN 47907, USA\\
\authormark{3} Department of Electrical and Electronic Engineering, Bangladesh University of Engineering and Technology, Dhaka 1205, Bangladesh\\
\authormark{4} Smuddin@uci.edu\\
\authormark{5} chowdh23@purdue.edu\\
\authormark{6} ehsan.kabir.buet@gmail.com\\}

\authormark{+}Authors contributed equally


\begin{abstract}
In this paper, we propose a surface plasmon resonance (SPR) structure based on Kretschmann
configuration incorporating layers of Silicon and BaTiO$_3$ on top of Ag for
real-time detection of severe acute respiratory syndrome coronavirus 2 (SARS-CoV-2)
using thiol tethered DNA 
as ligand. 
Extensive numerical analysis based
on transfer matrix theory as well as finite-difference time-domain (FDTD) technique has been performed to characterize the sensor response considering sensitvity, full width at half maxima and minimum reflection.
About 7.6 times enhanced sensitivity has been obtained using the proposed architecture for SARS-CoV-2 detection, compared to the basic Kretschmann
configuration. Notably, the structure provides consistent enhancement over other competitive SPR structures for both angular and wavelength interrogations with a figure-of-merit of 692.28. Additionally, we repeated simulations for various ligate-ligand pairs to assess the range of applicability and robust performance improvement has been observed. As a result, the proposed sensor design provides a suitable configuration for highly sensitive, rapid, non-invasive 
biosensing which can be useful if adopted in experimental sensing protocols. 
\end{abstract}


\section{Introduction}
Real-time detection of severe acute respiratory syndrome coronavirus 2 (SARS-CoV-2) with high sensitivity in a non-invasive manner is a critical research arena during this pandemic of Coronavirus disease 2019 (COVID-19) \cite{bhalla2020opportunities}. Since its emergence in December, 2019 in Wuhan, China, the devastating rapid transmission and havoc caused by this virus has resulted in the World Health Organization (WHO) classifying this outbreak as a pandemic on March 12, 2020 \cite{whoreport}. As per the report on Aug 6, 2020, no.~of confirmed cases has soared upto 18,614,177, with a death toll of 702,642 \cite{whosite}. With no proven medication or vaccines in sight still for COVID-19, significant research endeavors have focused on early diagnosis and management to control the damage \cite{diagnosing}. The laboratory standard detection scheme is reverse transcriptase quantitative polymerase chain reaction (RT-qPCR), which attempts to identify the presence of nucleic acid-based genetic sequence of this virus \cite{diagnosing}. However, RT-qPCR takes at least 3 h and requires sophisticated testing facilities \cite{fet} in addition to suffering from false positives. To cope with the rapid spread of COVID-19, highly sensitive diagnostic methods without requiring extensive lab setup is of utmost importance. Another promising diagnostic method is Computed Tomography (CT) scan where multiple chest X-rays are captured across different angles to obtain three-dimensional (3D) images. But the challenge using such technique lies in being able to differentiate symptoms from other lung disorders, failure in doing so results in rather low specificity of around 25\% \cite{CT} in practice. Apart from these, other methods include biochemical tests targeting the viral protein or the antibodies and contact tracing \cite{bhalla2020opportunities}. Recently, Seo \textit{et al.~}proposed a field-effect transistor based sensor topology for SARS-CoV-2 detection from human nasopharyngeal swabs \cite{fet}. While such sensing mechanisms are being explored, there is still an urgent demand for ultra-sensitive, real-time, point-of-care diagnostic methods to further aid the practitioners in fighting against COVID-19. To that end, one potential domain is the usage of optical biosensors\cite{kurihara2002theoretical,chowdhury2016real,chowdhury2019robust}. They are preferable due to high selectivity and sensitivity, rapid response time, cost-effectiveness and multiplexing capabilities. 

In terms of optical biosensors, plasmonics is an attractive field with real-time and label-free detection features. Commercially available surface plasmon resonance (SPR) and localized surface plasmon resonance (LSPR) sensors have been used for disease \cite{chowdhury2017detection}  as well as viral strains detection such as SARS, MERS and influenza \cite{bhalla2020opportunities}. Taking into account the promise of plasmonic sensors for detection purposes, Qiu \textit{et al.~}\cite{lspr-cov} have developed a biosensor combining plasmonic photothermal (PPT) effect and LSPR for COVID-19 diagnosis. They utilize DNA receptors which selectively bond with RNA sequences from the SARS-CoV2 through nucleic acid hybridization. One of the key challenges of accurate detection using such plasmonic sensors is ensuring ultra-sensitive performance with excellent limit of detection. Along those lines, SPR sensing is the most mature and established technology providing very good resolution of detection. However, it is pivotal to numerically analyze the suitability of deploying such SPR sensors for detecting SARS-CoV2 and a proper sensor configuration is required to achieve desired level of performance. This motivates our study in this paper where we propose an SPR sensor configuration aiming towards SARS-CoV2 detection with high sensitivity and figure-of-merit. To enhance the overall sensor performance, we integrate layers of Si and BaTiO$_3$ on top of the metal layer (Ag). Thiol-tethered DNA has been used as receptor for sensing. The proposed sensing pipeline assumes collection of samples from human nasopharyngeal swabs and passing them through the sensing channel in a liquid solution while the sensor is illuminated from different angles. When the binding occurs between immobilized receptor on sensor surface and target biomolecules from the flowing sample, a shift in SPR angle  is observed the extent of which determines the sensitivity. Our transfer matrix based analysis corroborated with finite-difference time-domain (FDTD) simulation shows that the proposed sensor architecture is able to obtain very high sensitivity with low minimum reflection (Rmin) and improved figure-of-merit (FOM) compared to other reported results. Furthermore, we verified the robustness of the sensor design through analyzing its performance for other ligand-ligate combinations. 

The rest of the paper is organized as follows- section 2 contains the design considerations and modeling schemes for different layers, then the optimization of the sensor configuration is described in section 3, we analyze the results in detail and perform comparative analysis with regards to other proposed sensors in section 4 and the paper is concluded in section 5.

\begin{figure}[h!]
\centering\includegraphics[width=10cm]{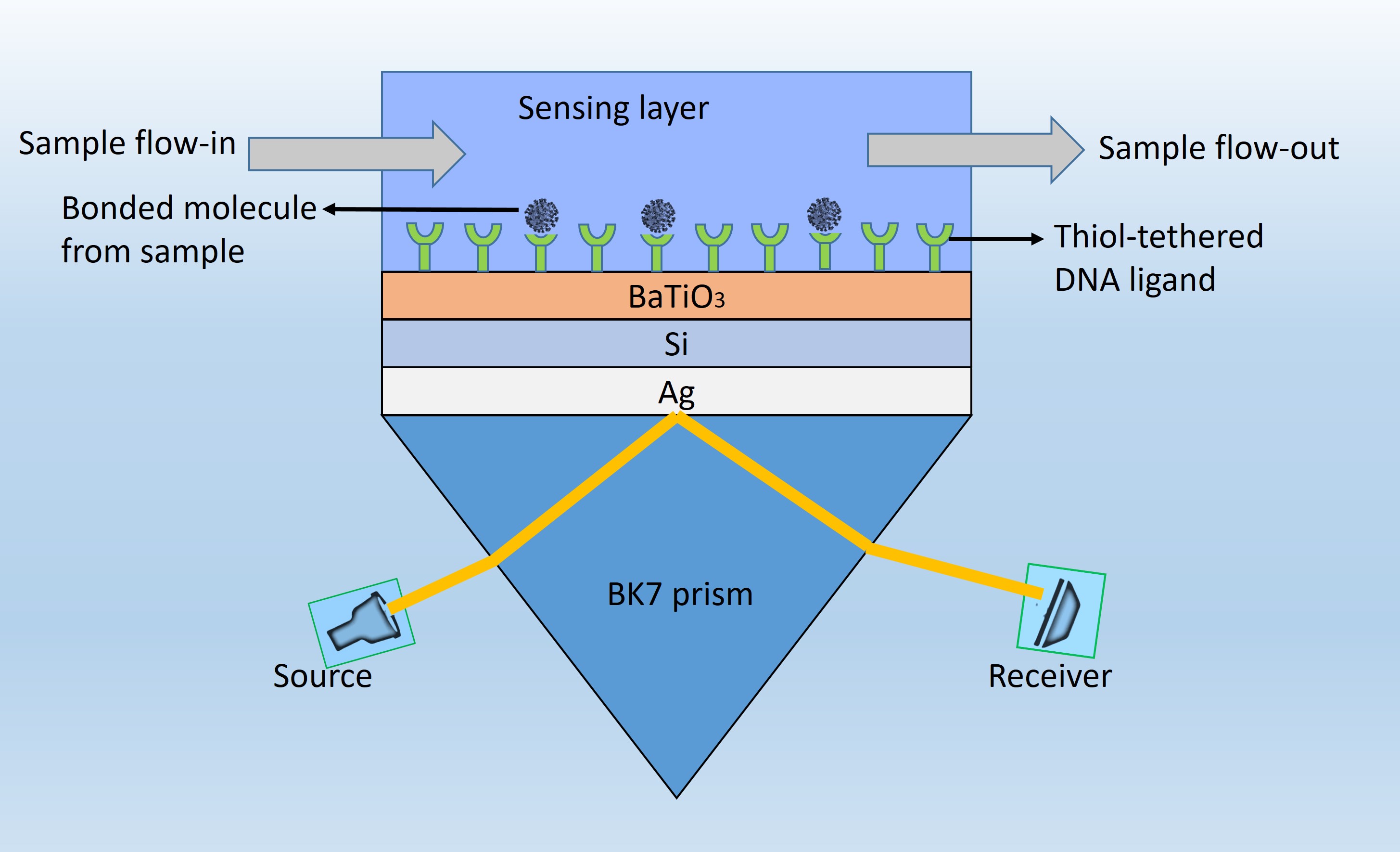}
\caption{Schematic illustration of the proposed structure}
\label{Fig1}
\end{figure}


\section{Design considerations and theoretical modeling of proposed structure}
The schematic view of our proposed high sensitivity SPR biosensor configuration is shown in Fig.~\ref{Fig1}. This novel hetero structure, as shown in the figure, is composed of 5 different layers. We use BK7 glass as a coupling prism followed by a Ag layer. Ag has been proven to demonstrate better sensitivity as a substrate layer due to higher SPR ratio \cite{Si4}. However, one of the major drawbacks of using Ag is its lesser chemical stability due to oxidation while interacting with bio molecules\cite{Si4,Ag1}. In our proposed design, we overcome this problem by employing 3 layers comprising of Si, BaTiO$_3$ and Thiol tethered DNA respectively, on top of Ag. Such configuration also aids in sensitivity enhancement. In recent research, Si has shown good promise in improving sensitivity  \cite{Si1}. Si's ability to enhance the electric field intensity of excitation light plays major role behind this sensitivity increment\cite{Si2,Si3}. BaTiO$_3$ also has desirable effects on sensitivity due to its high dielectric constant. Together with low dielectric loss, it has potential to contribute significantly to enhance sensitivity of our proposed structure \cite{BaTiO3}. Thiol tethered DNA is used as a ligand layer for sensing medium as it has shown excellent properties as a receptor of SARS-Cov-2\cite{lspr-cov,Thiol2}. Having decided on the composition of various layers of our sensor design, the next step is to model them properly to simulate their response. We model each layer as a homogeneous continuous medium upto the next layer interface wherein the layer is represented by its refractive index. Hence, the next subsection focuses on modeling the refractive indices of the layers under consideration.
\subsection{Modeling of layers}
The refractive indices of different layers of our structure are each calculated separately. The first layer is a coupling prism of BK7 glass. The refractive index of BK7 glass can be calculated from the following relation\cite{Bk7}:  
\begin{equation}
    n_1 = (\frac{1.03961212\lambda^2}{\lambda^2 - 0.00600069867} + \frac{0.231792344\lambda^2}{\lambda^2 - 0.0200179144} + \frac{1.01046945\lambda^2}{\lambda^2 - 103.560653} + 1 )^\frac{1}{2} ,
\end{equation}
where $\lambda$ is the wavelength of incident light in micrometer. This equation is only applicable for wavelength between 0.37 to 2.5 $um$\cite{BK7wave}. The refractive index of Ag layer can be defined using the well-known drude model for metal\cite{AgRefr},
\begin{equation}
     n_m = ( 1 - \frac{\lambda^2\lambda_c}{{\lambda_p}^2(\lambda_c + i\lambda)} )^\frac{1}{2} ,
\end{equation}
where $\lambda_p = 1.4541\times10^{-7}m$ and $\lambda_c = 1.7614\times10^{-5}m$ are plasma and collision wavelength for Ag, respectively. The next layer of our proposed structure is silicon, the refractive index of which is defined as \cite{Si3}:
\begin{equation}
    n_{Si} = A_1 + A_2e^{-\frac{\lambda}{t_0}} + A_3e^{-\frac{\lambda}{t_1}},
\end{equation}
where $\lambda$ is the wavelength of the incident light in um, A$_1$ = 3.44904, A$_2$ = 2271.88813, A$_3$ = 3.39538, $t_0 = 0.058304$ and $t_1 = 0.30384$. The experimentally obtained refractive index data for BaTiO$_3$  at our operating wavelength is 2.4043\cite{BaTO3refr}. Finally, thickness dependent refractive index data of Thiol tethered DNA is obtained from experimental results\cite{Thiolrefr}. With the proper models of various layers, we now need to  characterize the optical response of the sensor for which suitable optical excitation through appropriate irradiation is required first. This begets our next discussion on  input lighting scheme.
\subsection{Illumination strategy}
The choice of operating wavelength needs to be well balanced between sensitivity and optical non-linearity. Sensitivity and optical non linearity both are proportionally varied with frequency (inversely proportional to wavelength) \cite{Si4}. Optical Kerr effect becomes prominent if the frequency of operation is increased such that the beam intensity is on the order of 1 GW $cm^{-2}$ and exerts significant variations in refractive index \cite{anower2018performance}. We select operating wavelength as 633 $nm$ such that sensitivity is enhanced with minimum possible Kerr effect \cite{Si4,anower2018performance}. We employ p-polarised light in Attenuated Total Internal Reflection (ATR) configuration in our analysis. In this case, the incident angle must be greater than the critical angle $\theta_c$ of glass-silver interface where, $\theta_c = \frac{n_{Ag}}{n_{Air}}$. If we assume energy loss other than material absorption to be zero, then according to principle of energy conservation, the total of absorption(A), transmission (T) and reflection(R) must be equal to 1 (i.e., A+T+R = 1). Again, since transmission is zero under ATR, absorption reduces to A = 1-R. When surface plasmons (SPs) are excited, reflection becomes minimum. As a result, this leads to the maximum transfer of light energy to the evanescent wave. With the structural and illumination setup discussed above, next it is important to formulate a theoretical framework to numerically evaluate the sensor performance, this leads to our following section on numerical modeling.
\subsection{Theoretical framework }
We have performed numerical analysis to determine the change of reflectivity using transfer matrix method \cite{TM} and Fresnel equation \cite{Fresnel} for basic Kretschmann configuration having multiple layers. Our proposed model consists of five layers with BK7 prism, silver, Silicon, Barium titanate and thiol tethered DNA being placed in parallel one after another. The thickness of each layer varies in perpendicular direction denoted as z-axis. The boundary condition at the interface of first two layers and the last two layers are considered as ${Z=Z_1=0}$ and ${Z=Z_{n-1}}$ respectively. The dielectric constant($\epsilon_p$) of p'th layer equals to the square of its refractive index ($n_p$). These methods adopt no approximation allowing them to give accurate results rapidly. The transfer matrix expresses a relationship among the tangential components of electric and magnetic fields of the first layer and the last layer, and the relation is given as\cite{st1}-
\begin{equation}
    \begin{bmatrix}
    E_1 \\ B_1
    \end{bmatrix}
    = M \times
     \begin{bmatrix}
    E_{n-1} \\ B_{n-1}
    \end{bmatrix},
\end{equation}
where, $E_1$ and $B_1$ are tangential components of electric and magnetic fields respectively in the interface of the first two layers, and $E_{n-1}$ and $B_{n-1}$ are the same for the interface of the N-1 and $N^{th}$ layer. $M$ represents the characteristic matrix of the whole structure on which the reflection coefficients are dependent. For TM polarized light it can be expressed as,
\begin{equation}
   M = \left(\prod_{k=2}^{N-1} M_k\right)=
   \begin{pmatrix}
        M_{11} & M_{12}\\ 
        M_{21} & M_{22}
    \end{pmatrix},
\end{equation}
where,
$M_k$ = 
$\begin{bmatrix}
        \cos{\delta_k} & (\frac{-i\sin{\delta_k}}{\gamma_k})\\ 
        (-i\gamma_k\times\sin{\delta_k})& \cos{\delta_k}
\end{bmatrix}$ \
with $\gamma_k$ = $\sqrt{\frac{\mu_k}{\epsilon_k}} \times \cos{\theta_k}$ =
$\frac{\sqrt{(\epsilon_k-{n^2_1}\sin^2{\theta_1})}}{\epsilon_k}$, and \ \ \ \ \ \ \ \ \ \ \
$\delta_k$ = $\frac{2\pi}{\lambda}n_k\cos\theta_k(Z_k-Z_{k-1})$ = $\frac{2\pi{d_k}}{\lambda} \sqrt{(\epsilon_k-{n^2_1}\sin^2{\theta_1})}$ \ \ \ in which $\mu_k$, ${\epsilon_k}$, $d_k$, $\theta_1$, $\lambda$ are magnetic permeability, electric permittivity, thickness of $k^{th}$ layer, incident angle and wavelength of light respectively. 
\hfill\\
\hfill\\
\hfill
From the above equations, we can derive that the total reflection coefficient($r_p$) as,
\begin{equation}
   r_p = \frac{(M_{11}+M_{12}\gamma_N)\gamma_1 - (M_{21}+M_{22}\gamma_N)}{(M_{11}+M_{12}\gamma_N)\gamma_1 + (M_{21}+M_{22}\gamma_N)},
\end{equation}
where, $M_{11}, M_{12}, M_{21}, M_{22}$ are the components of the transfer matrix. The reflectivity ($R_p$) for a TM polarized light is calculated as ${\lvert {r_p} \rvert}^2$. 

We also utilize FDTD method to analyze our structure in order to further strengthen the validity of our investigation. In FDTD, we consider meshing so that it can reach convergence with sufficient precision without making the simulation unnecessarily lengthy. It simulates each layer with corresponding complex refractive index at the respective wavelength. We apply perfectly matched layer (PML) boundary condition along the direction of light incidence and bloch boundary condition in the direction perpendicular to it. With the structural setup and numerical framework as discussed above at hand, next we need to define some standard evaluation criteria to quantify the performance of our design which leads to our next section.  
\subsection{Performance parameters}
To evaluate the performance of SPR biosensor, several criteria can be utilized such as Sensitivity (S), Minimum reflectance (Rmin) and full width at half maximum (FWHM). These parameters can be obtained from the curve of incident angle versus reflection. To analyze the sensing performance and compare with other novel structures, we defined Figure of Merit (FOM) as follows,
\begin{equation}
    FOM = \frac{S}{FWHM\times Rmin}.
\end{equation}
Minimum reflectance is defined as the reflectance at the surface plasmon resonance angle. To increase the detection accuracy, the value of the Rmin should be as low as possible. Resolution and signal to noise ratio (SNR) are dependent on FWHM. FWHM is characterized as the spectral width of SPR curve at half maxima. The most crucial parameter is the sensitivity of the sensor which is the characteristic parameter of a sensor that transforms a device into a sensor. It can be observed by monitoring the shift of the resonance angle of SPR curve due to the change in the refractive index of the sensing medium. Quantitively, it can be calculated from the following equation,
\begin{equation}
    S = \frac{{\delta\theta}_{SPR}}{\delta n_{res}},
\end{equation}
where $\theta_{SPR}$ is the SPR angle and $n_{res}$ is the refractive index of the corresponding target molecule. While sensitivity determies the characteristic response of a sensor, the true performance of it cannot be evaluated on sensitivity  solely. Therefore, for a more comprehensive analysis, we consider FOM taking into account the overall effect of S, FWHM and Rmin as mentioned  before.

\section{Optimization of the structure}
As mentioned earlier, there are 5 different layers in our proposed architecture, each layer has significant impact on the overall sensor performance. Therefore, to harness the maximum response out of our sensor configuration, proper optimization of design parameters needs to be performed in a synergistic manner. Here, we follow layerwise optimization procedure as outlined in \cite{kabir2020optimization}. First, we investigate sequentially the effects of each layer thickness on different performance parameters of the sensor. For this process, we determine the reflectivity as a function of incident angle using Eqn. 6. The minimum value of the reflectivity is termed as Rmin. We also calculate sensitivity as a function of incident angle utilizing Eqn. 8. Our goal is to make Rmin as low as possible while keeping the sensitivity as high as possible. For this purpose, each layer thickness is optimized one at a time keeping the thickness of all other layers fixed. Initially, we set the thickness of Si, BaTiO$_3$, Thiol layers at some arbitrary values and optimize Ag layer thickness to get minimum Rmin and maximum sensitivity. Then the same process is repeated for other layers as well based on sensitivity and Rmin. After that, we optimize the Ag thickness again with optimized thickness of Si, $BaTiO_3$ and Thiol thickness. This optimization process is continued in an iterative manner until we get the most optimized structure possible.  In Fig.~2(a), we show dependence of Rmin and sensitivity of the sensor against the thickness variation of Ag. Rmin gradually decreases with increasing thickness up to a certain thickness and then it starts increasing. Similar trend holds true for sensitivity as well but in opposite direction. We see that optimum thicknesses for minimum Rmin and maximum sensitivity are very close. Based on the obtained results, we select 50 $nm$ as Ag thickness for our structure. Notably, our numerically optimized metal thickness matches well with theoretically obtained optimized value for metal layer as described in \cite{uddin2017theoretical}. Figures 2(b), (c) and (d) show the effect of Si, BaTiO$_3$ and thiol layer thickness variations on Rmin and sensitivity, respectively. For Si, the variation pattern follows more or less same fashion as Ag. After 8 $nm$ there is a drastic increase in Rmin. So, we choose 8 $nm$ for Si thickness which gives maximum sensitivity with minimal Rmin.  For $BaTiO_3$, we select 11 $nm$ as thickness for the same reason. Thiol optimization curve is quite different. Rmin was very low and fairly constant until 3 $nm$ of thiol thickness, after which it starts increasing exponentially. However, the sensitivity peaks just before 4 $nm$. As a result, we opt for 4 $nm$ as thiol thickness to attain the best compromise between low Rmin and high sensitivity in our proposed structure.
\begin{figure}[h!]
\begin{subfigure}{.5\linewidth}
\centering\includegraphics[width=2 in]{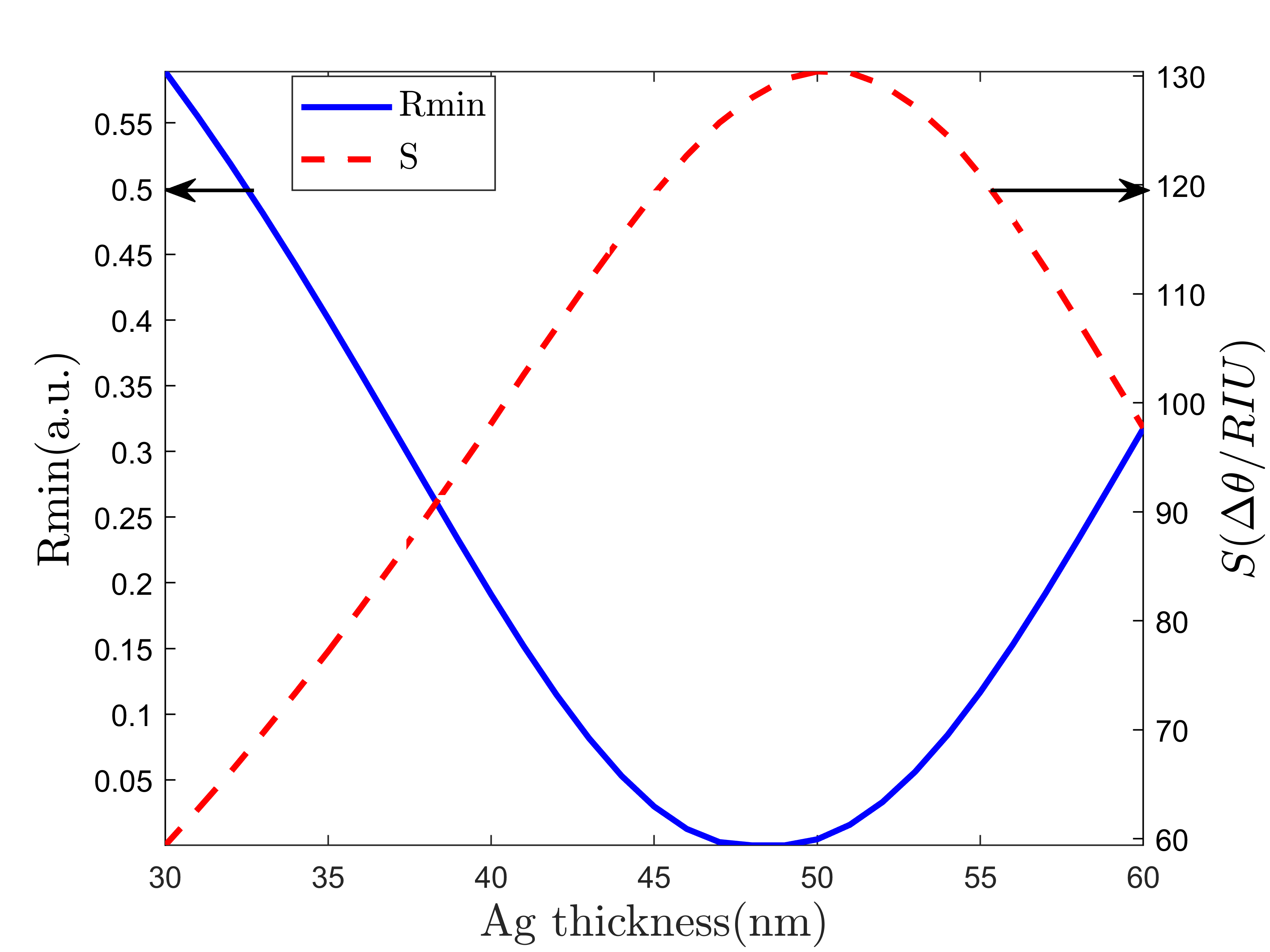}
\subcaption{Ag thickness Variation}
\label{opt1}
\end{subfigure}
\begin{subfigure}{.5\linewidth}
\centering\includegraphics[width=2 in]{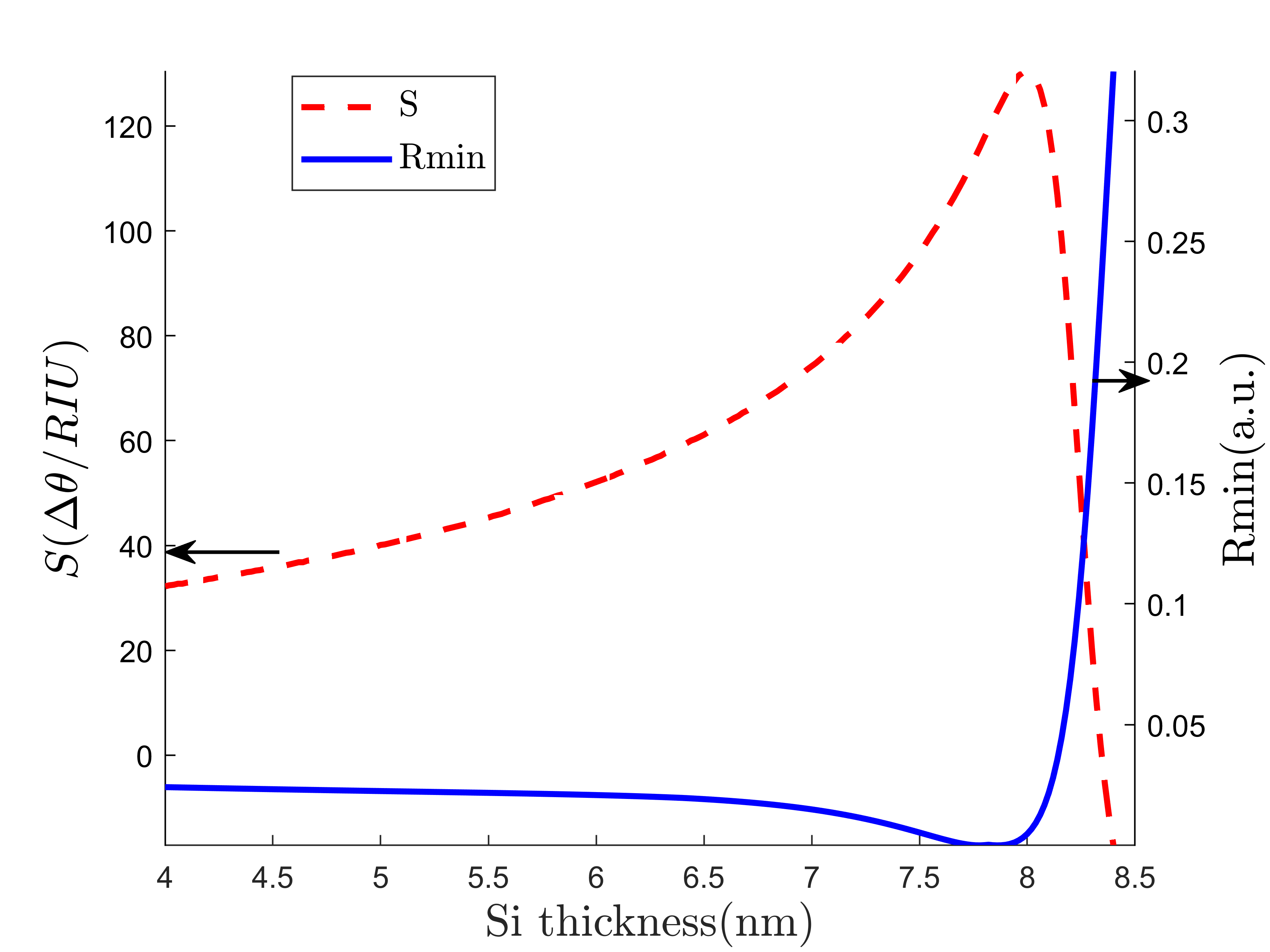}
\subcaption{Si thickness Variation}
\label{opt2}
\end{subfigure}
\begin{subfigure}{.5\linewidth}
\centering\includegraphics[width=2 in]{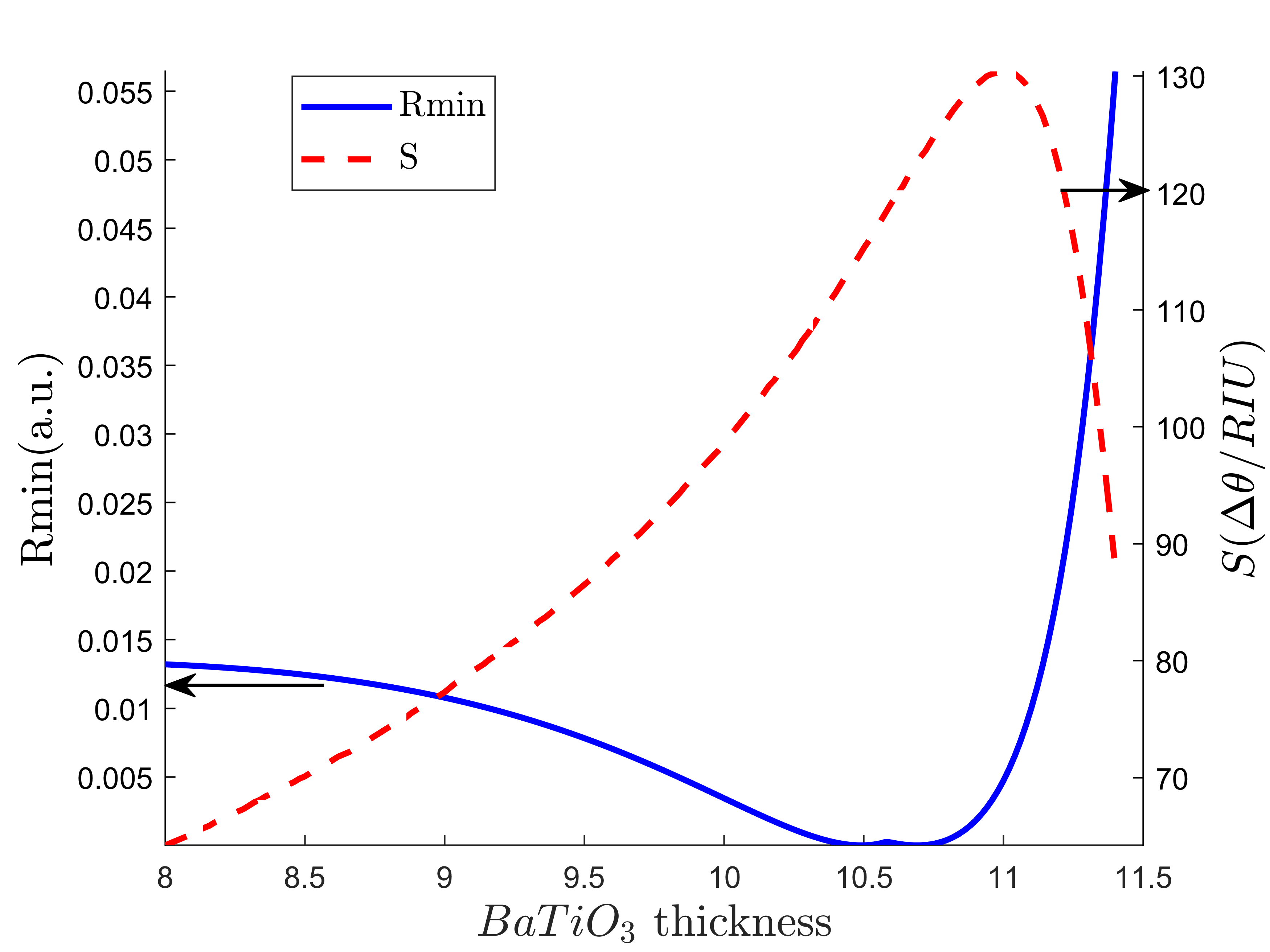}
\subcaption{$BaTiO_3$ thickness Variation}
\label{opt3}
\end{subfigure}
\begin{subfigure}{.5\linewidth}
\centering\includegraphics[width=2 in]{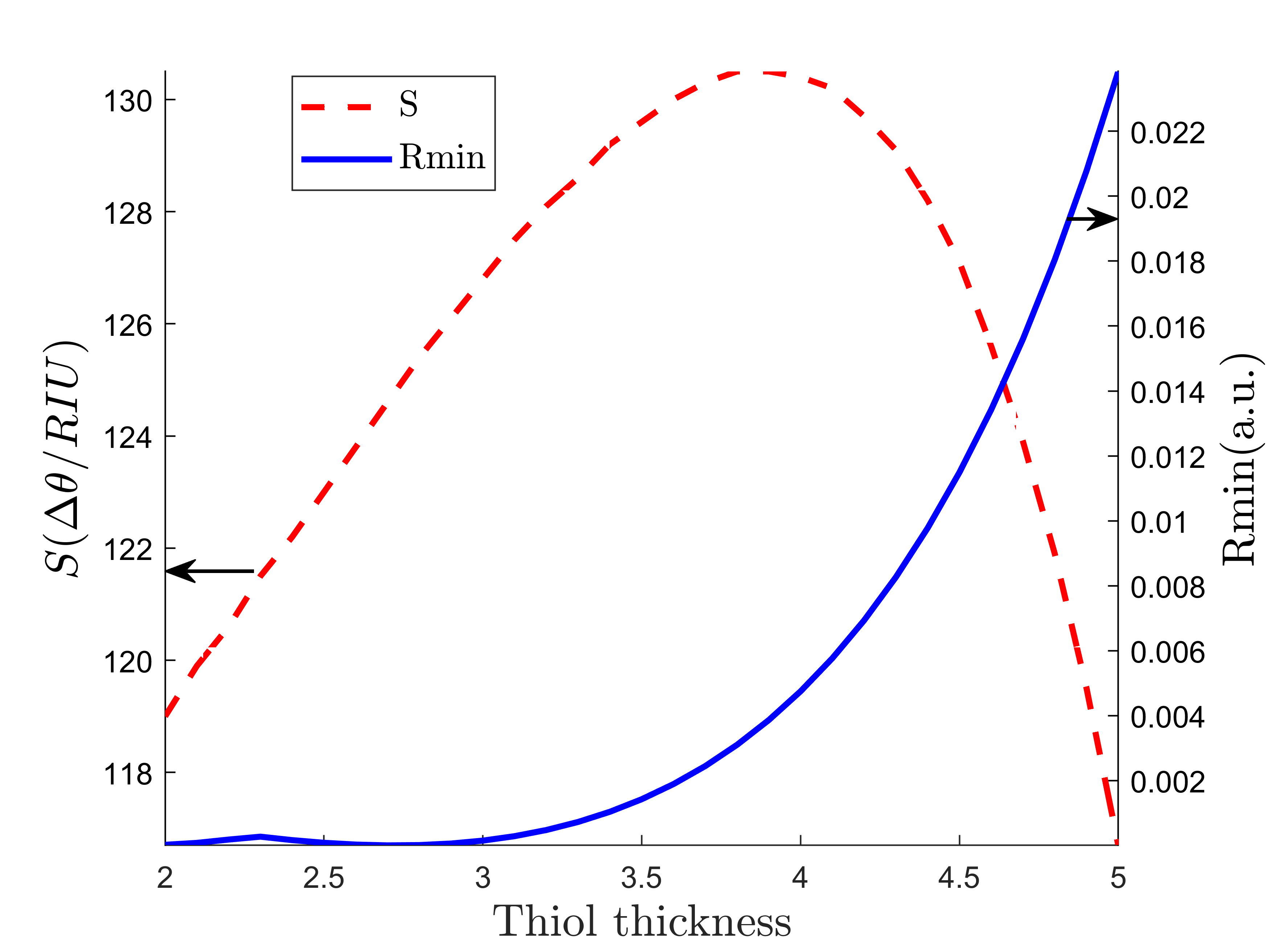}
\subcaption{Thiol tethered DNA thickness Variation}
\label{opt4}
\end{subfigure}
\caption{Effect of different layer thickness on Rmin and sensitivity}
\label{opt}
\end{figure}
\section{Results and discussions}

As mentioned in previous sections, the structure presented in this work consists of 5 layers. The fifth and final layer is the sensing layer the refractive index($n_s$) of which will vary based on the sample attachment. These variations lead to the change in the resonance angle of SPR curve. From such variations, we obtain the sensitivity of the sensor according to Eqn.~8. Though sensitivity is the characteristic parameter for SPR bio-sensor's performance, other parameters such as Rmin, FWHM etc. have significant effect on the overall detection accuracy. Here, we will analyze our design based on different performance parameters and compare with other novel structures. We will also study the robustness of our sensor with different ligand-ligate pairs.
\subsection{Performance analysis}
\begin{figure}[h!]
\begin{subfigure}{.5\linewidth}
\centering\includegraphics[width=2 in]{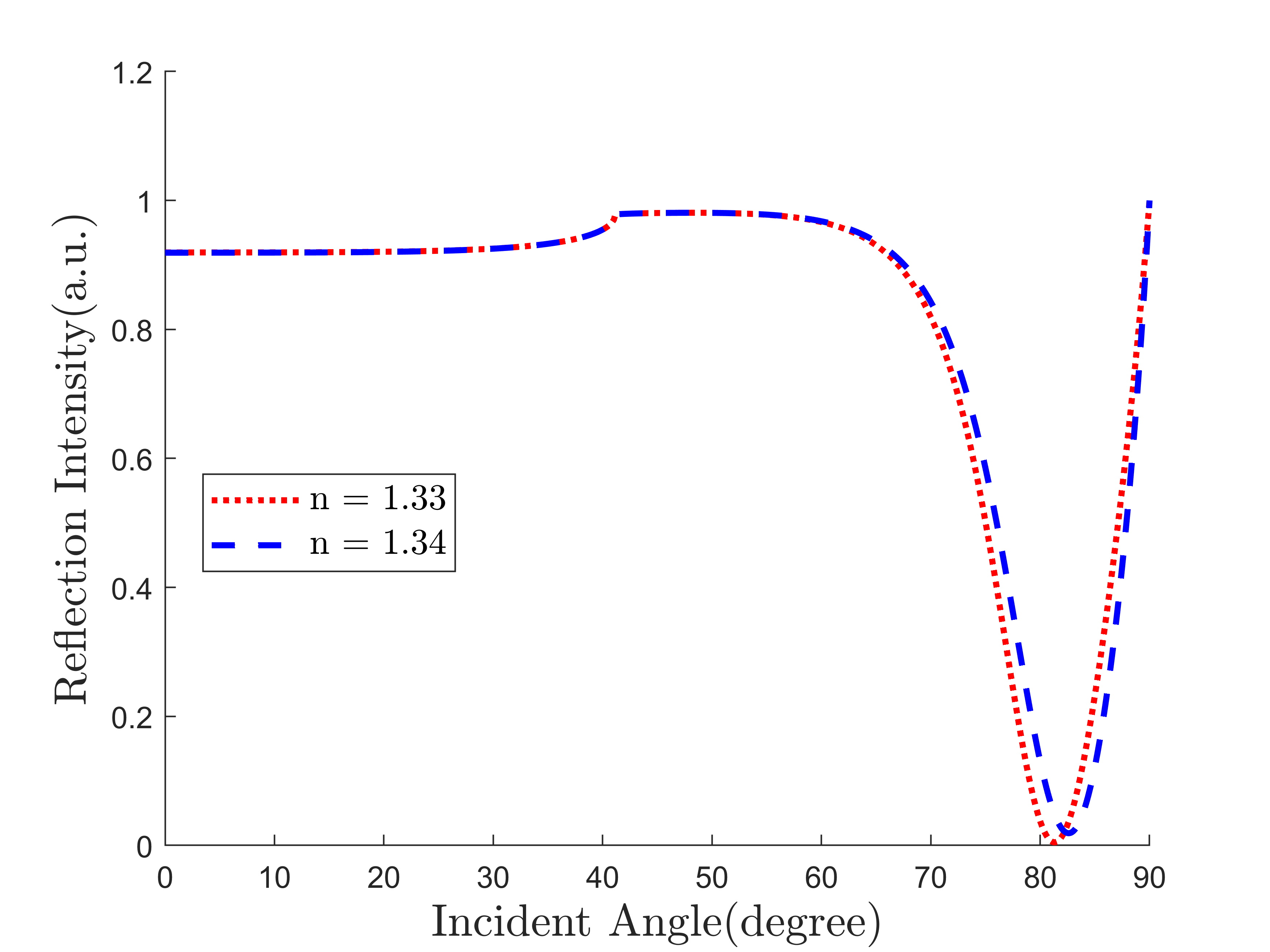}
\subcaption{Reflected light intensity versus source incident angle}
\label{main1}
\end{subfigure}
\begin{subfigure}{.5\linewidth}
\centering\includegraphics[width=2 in]{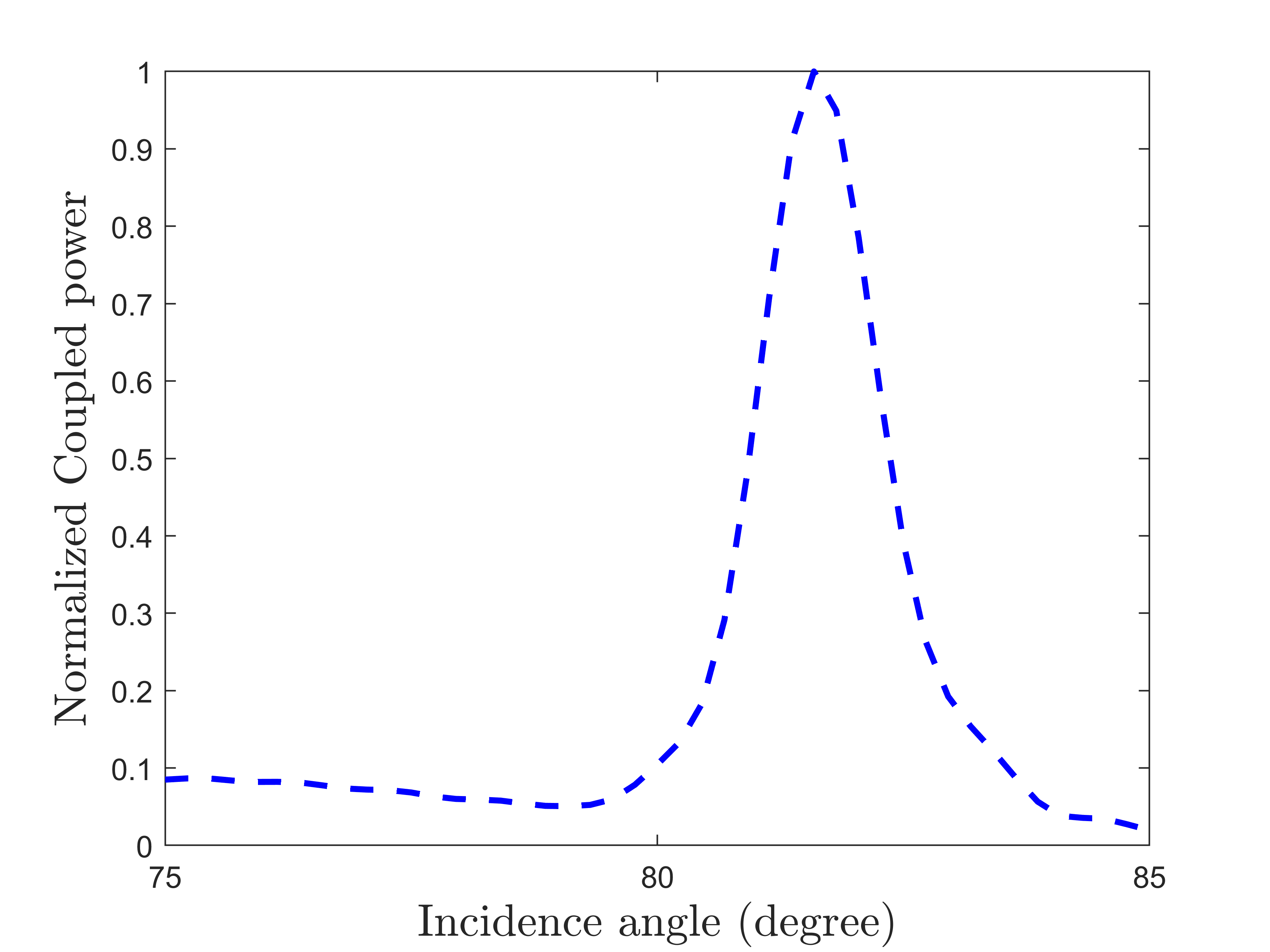}
\subcaption{Normalised coupled power versus source incident angle }
\label{main2}
\end{subfigure}
\begin{subfigure}{.5\linewidth}
\centering\includegraphics[width=2 in]{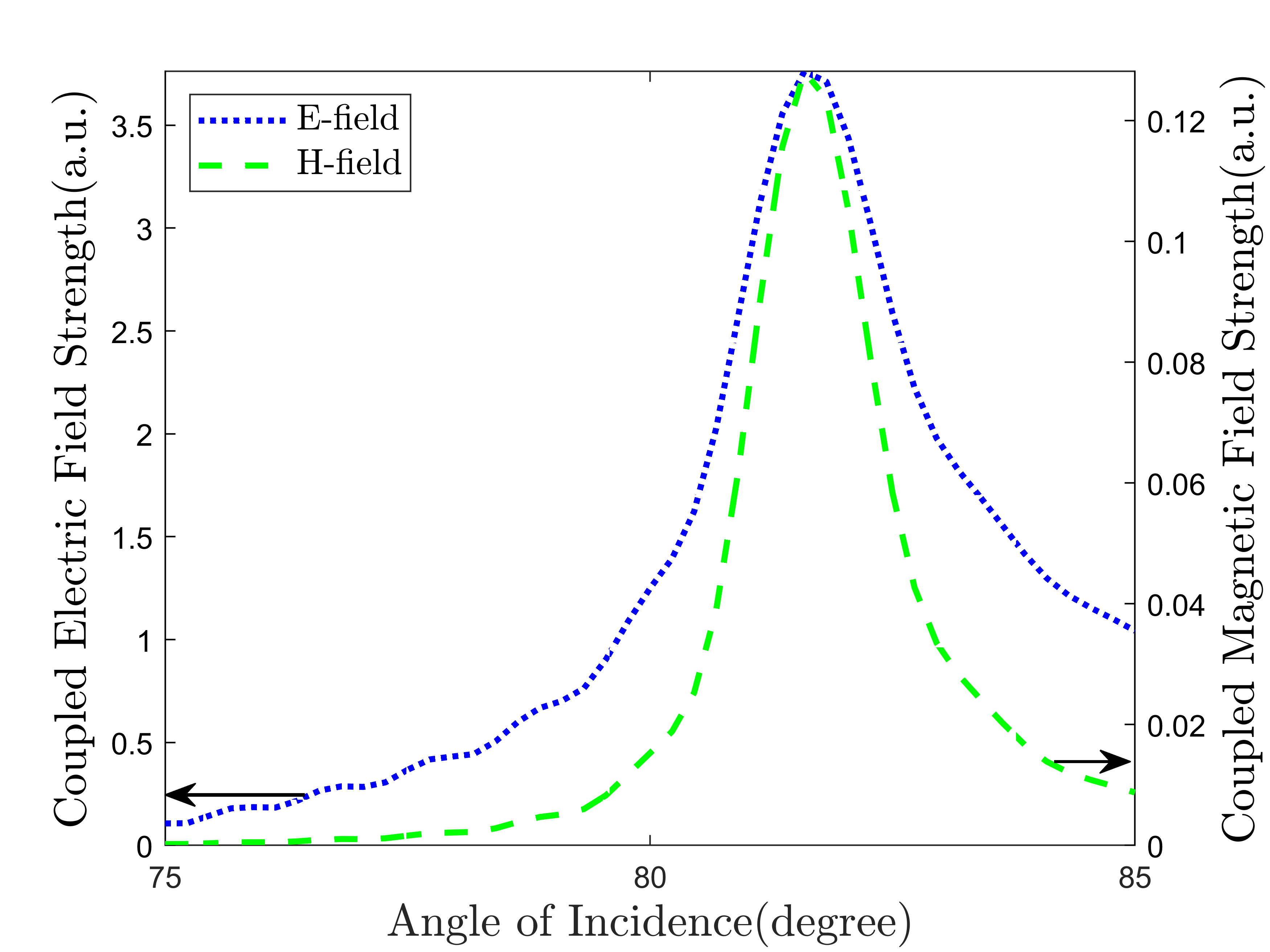}
\subcaption{Coupled field strength versus source incident angle}
\label{main3}
\end{subfigure}
\begin{subfigure}{.5\linewidth}
\centering\includegraphics[width=2 in]{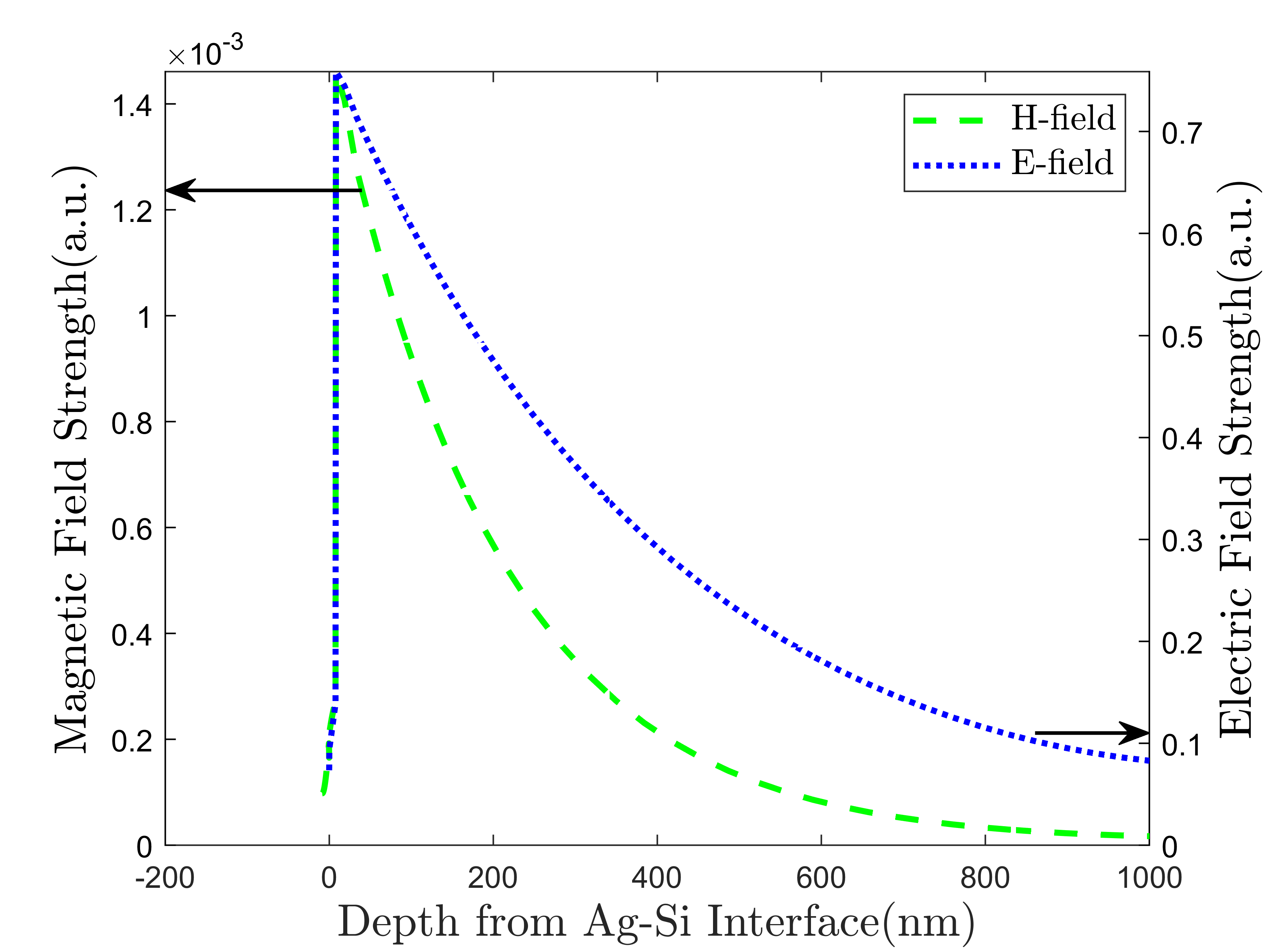}
\subcaption{Optical near field distribution of the plasmonic films}
\label{main4}
\end{subfigure}
\caption{Performance analysis of the proposed structure}
\label{main}
\end{figure}

The performance of a sensor is determined by sensing accuracy as well resolution. For SPR biosensor, resolution depends on FWHM of the reflectivity curve. The narrower the curve and hence lower FWHM, the better the resolution is. On the other hand, sensing accuracy depends on Rmin and sensitivity of the device. In Fig.~\ref{main1} we show the SPR curve for two different refractive index of the sensing layer. We see that for a slight increase of refractive index, there is a noticeable variation in the resonance angle. This gives a sensitivity as high as 130.3 deg/RIU which is 7.6 times higher than the conventional Kretschmann configuration. In Fig.~\ref{main2} and Fig.~\ref{main3}, we plot the coupled power and coupled fields versus incident angle, respectively. We see that both the coupled electric and magnetic field are maximum at the resonance angle. Same thing happens to the couple power as well which confirms the SPR occurrence at that particular incident angle. This clearly indicates the creation of evanescent wave at this specific angle and consequently we get a sharp dip in resonance curve. When light is incident at the resonance angle, maximum power transfer occurs from the source to the evanescent wave. Electric and 
magnetic field profiles are shown in Fig.~\ref{main4} along the direction of penetration of evanescent wave inside the sensing layer. Both the E and H-fields follow  almost the same pattern. Field intensity peaks sharply at the vicinity of metal-dielectric interface and penetrates through the dielectric layers. As it passes through the dielectric layers, it dies out exponentially. Essentially, field strength is very strong within a very small distance where the analyte bonds with the ligand. This indicates a very narrow range of detectable distance. In our case, analyte bonding starts after 23 $nm$ from the interface. We see that fields are very strong in this region which provides the enhanced sensitivity.

We analyze all the structures from the basic Kretschmann configuration to our proposed design through subsequent intermediate structures. Gradual development of performance is noticed as we move from basic to proposed design. We vary refractive index from 1.33 to 1.34 for the sensing layer and observe the corresponding change in the
\begin{figure}[h!]
\begin{subfigure}{.5\linewidth}
\centering\includegraphics[width=2 in]{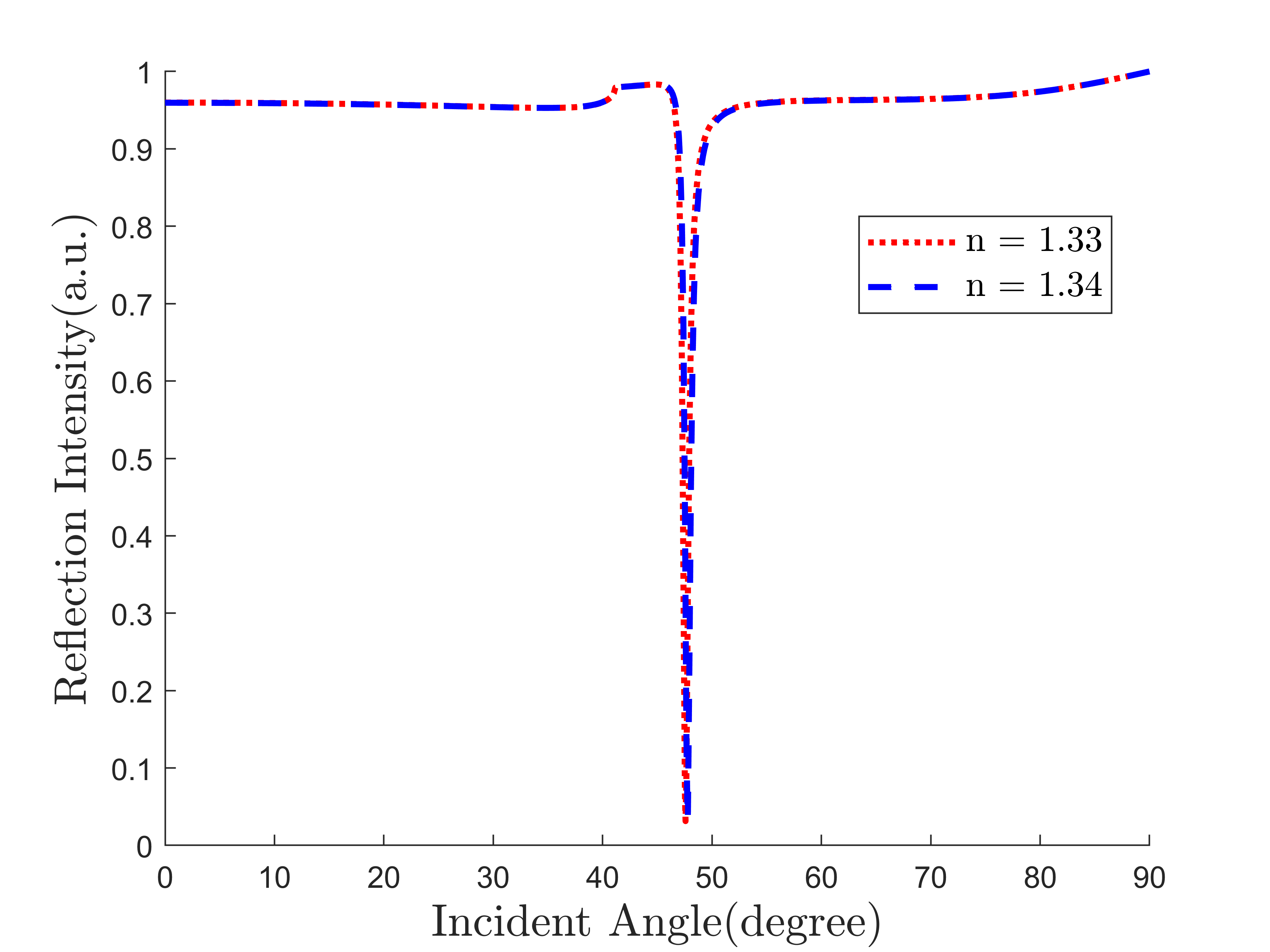}
\subcaption{Basic configuration}
\label{dev1}
\end{subfigure}
\begin{subfigure}{.5\linewidth}
\centering\includegraphics[width=2 in]{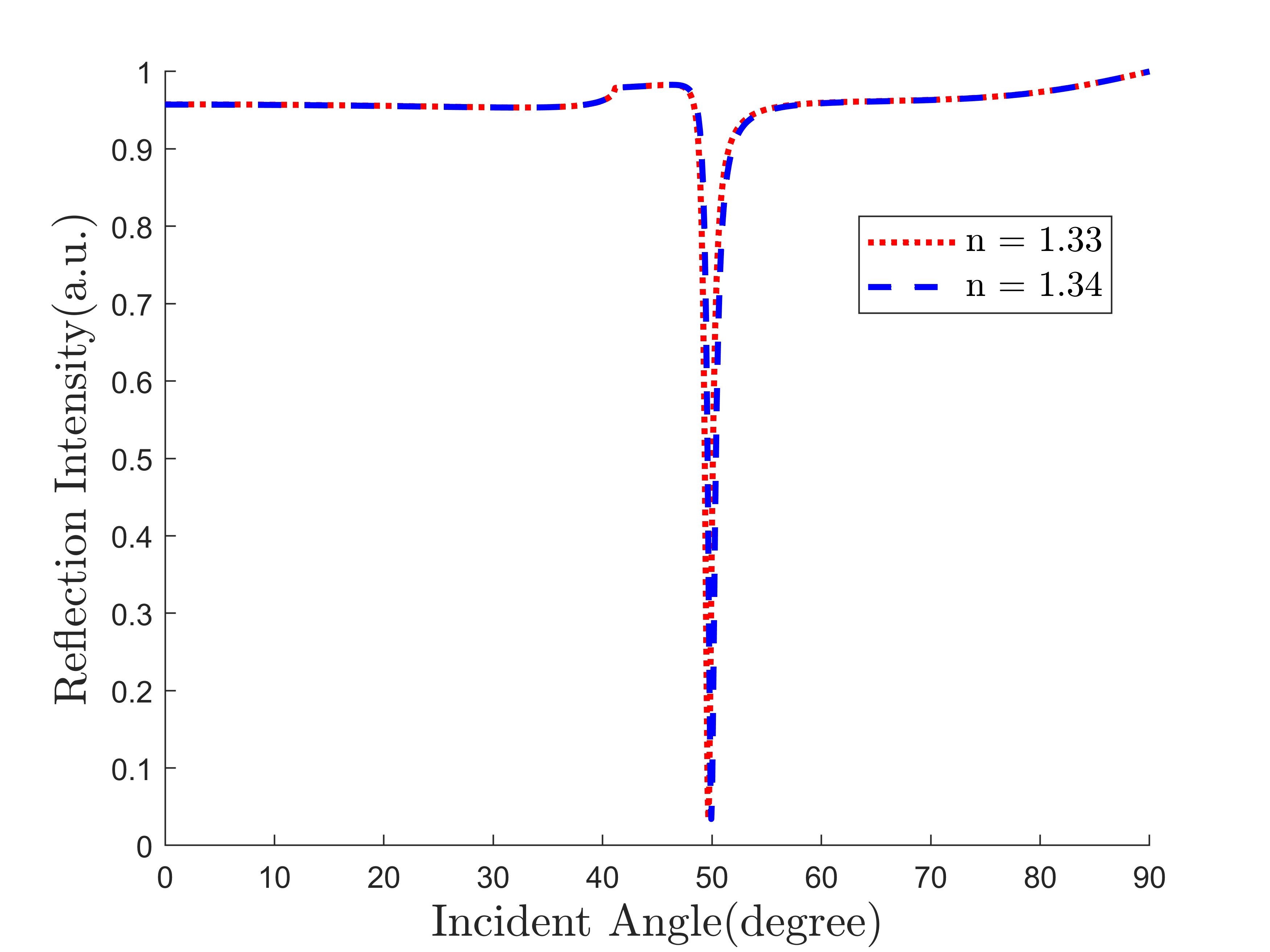}
\subcaption{Basic configuration with Thiol tethered DNA }
\label{dev2}
\end{subfigure}
\begin{subfigure}{.5\linewidth}
\centering\includegraphics[width=2 in]{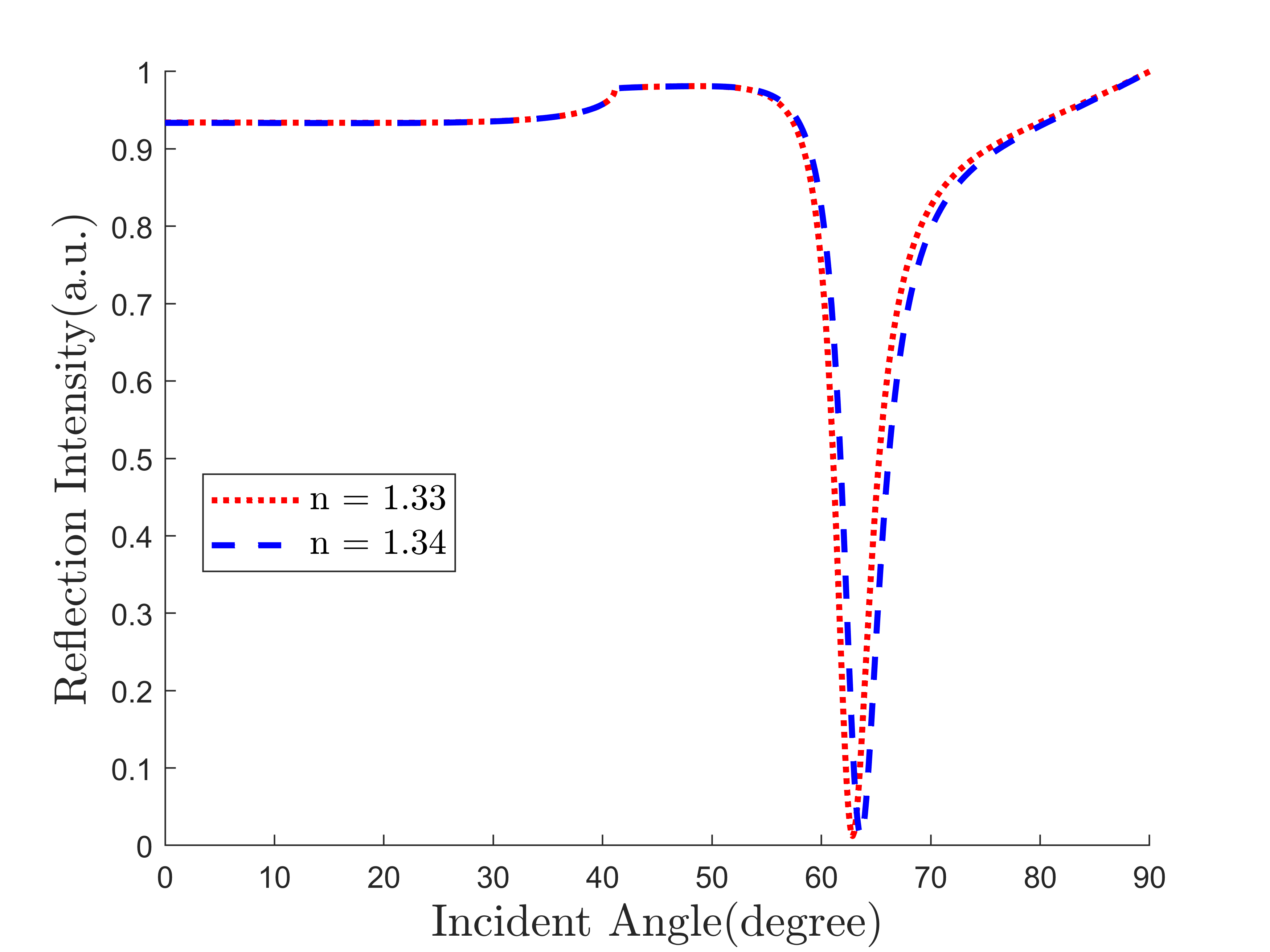}
\subcaption{Ag-Si-Thiol tethered DNA}
\label{dev3}
\end{subfigure}
\begin{subfigure}{.5\linewidth}
\centering\includegraphics[width=2 in]{Angle_Interrogation_main.png}
\subcaption{Proposed Structure}
\label{dev4}
\end{subfigure}
\caption{Reflectance versus incident angle}
\label{dev}
\end{figure}
resonance angle. As we see in the Fig.~\ref{dev}, sensitivity is significantly increased for the proposed structure. For the basic structure, we get the sensitivity to be only 18.4 degree/RIU. As we need a suitable receptor for Covid-19 sample, we put Thiol tethered DNA on top of Ag layer which increases sensitivity a little bit to 26.7 degree/RIU. 
\begin{table}[h!]
\centering
\fontsize{9}{9}\selectfont
\raisebox{0.08\height}{
\parbox{.5\linewidth}{
\caption{Different Layers' Effect on Sensitivity}
\begin{tabular}{|c|c|}
\hline
\textbf{\begin{tabular}[c]{@{}c@{}}Sensor\\ Configuration\end{tabular}} & \textbf{Sensitivity (degree/RIU)} \\ \hline
Basic               & 18.4  \\ \hline
Basic+Thiol         & 26.7  \\ \hline
Basic+Silicon       & 71.9  \\ \hline
Basic+Silicon+Thiol & 75.2  \\ \hline
Proposed architecture              & 130.4 \\ \hline
\end{tabular}
}}
\end{table}
Inclusion of Si layer in between Ag and Thio tethered DNA increases the sensitivity almost 3 times to 75.2 degree/RIU. Finally for our proposed structure, we find sensitivity almost doubled to 130.3 degree/RIU. It is  about 7.6 times compared to the basic configuration. This analysis is summarized in Table 1. In Fig.~\ref{Fieldvdepth}, we plot field strength against depth from the Ag-Si interface  along the direction of evanescent wave penetration (z-axis in our case). We see that the peak E-field undergoes a slight right shift with added layers. The rightmost red curve is our proposed structure. As we have seen earlier, our proposed structure contains some extra layers which shift the molecular binding region towards right further. With the peak E-field shifted to the right, molecules now bond in the significant electric field region. This happens for all intermediate structures. 
\begin{figure}[h!]
\centering\includegraphics[width= 2 in]{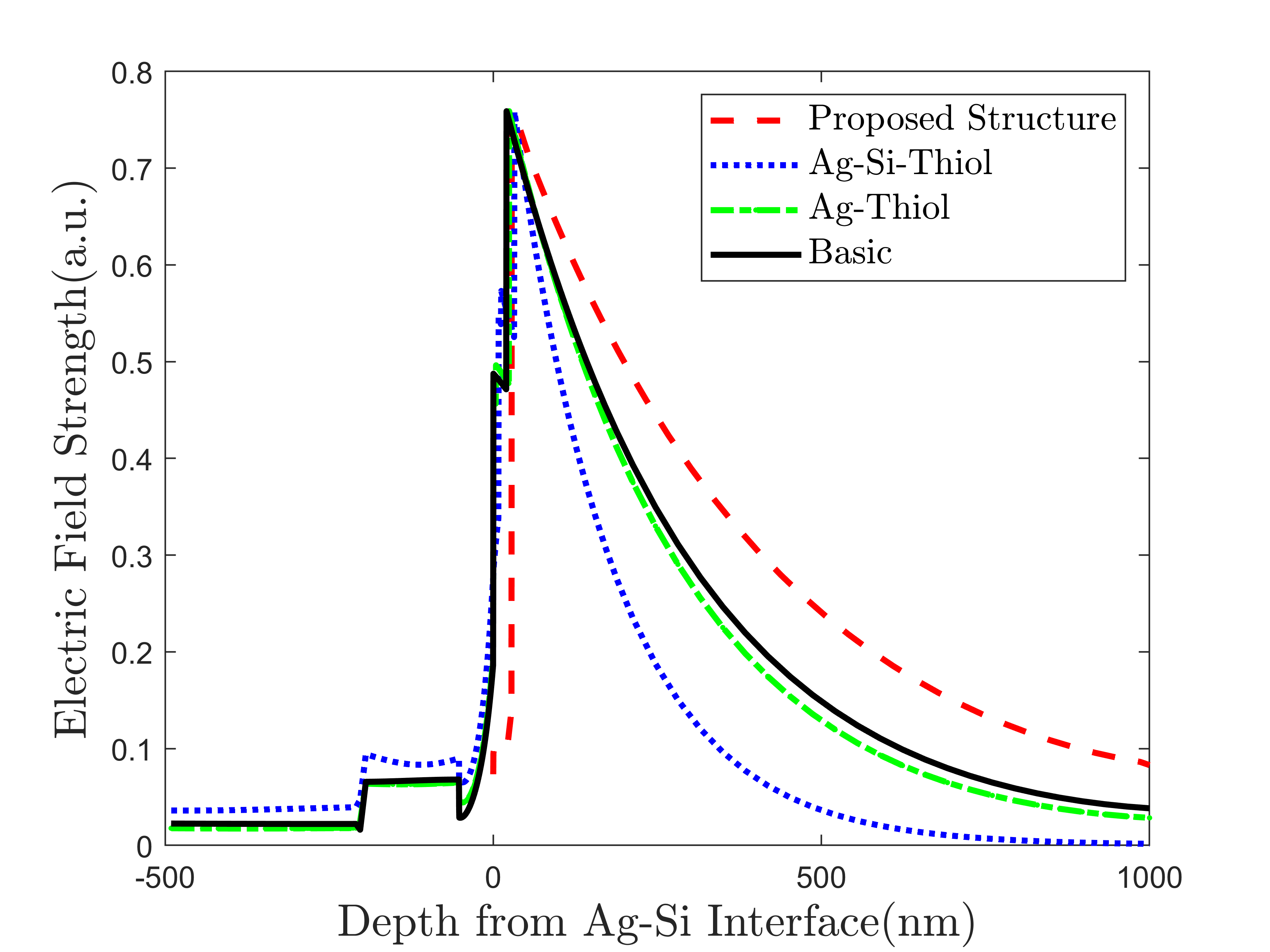}
\caption{Evolution of optical near field distribution of the plasmonic films for different
structures}
\label{Fieldvdepth}
\end{figure}

\begin{figure}[h!]
\centering\includegraphics[width= 2 in]{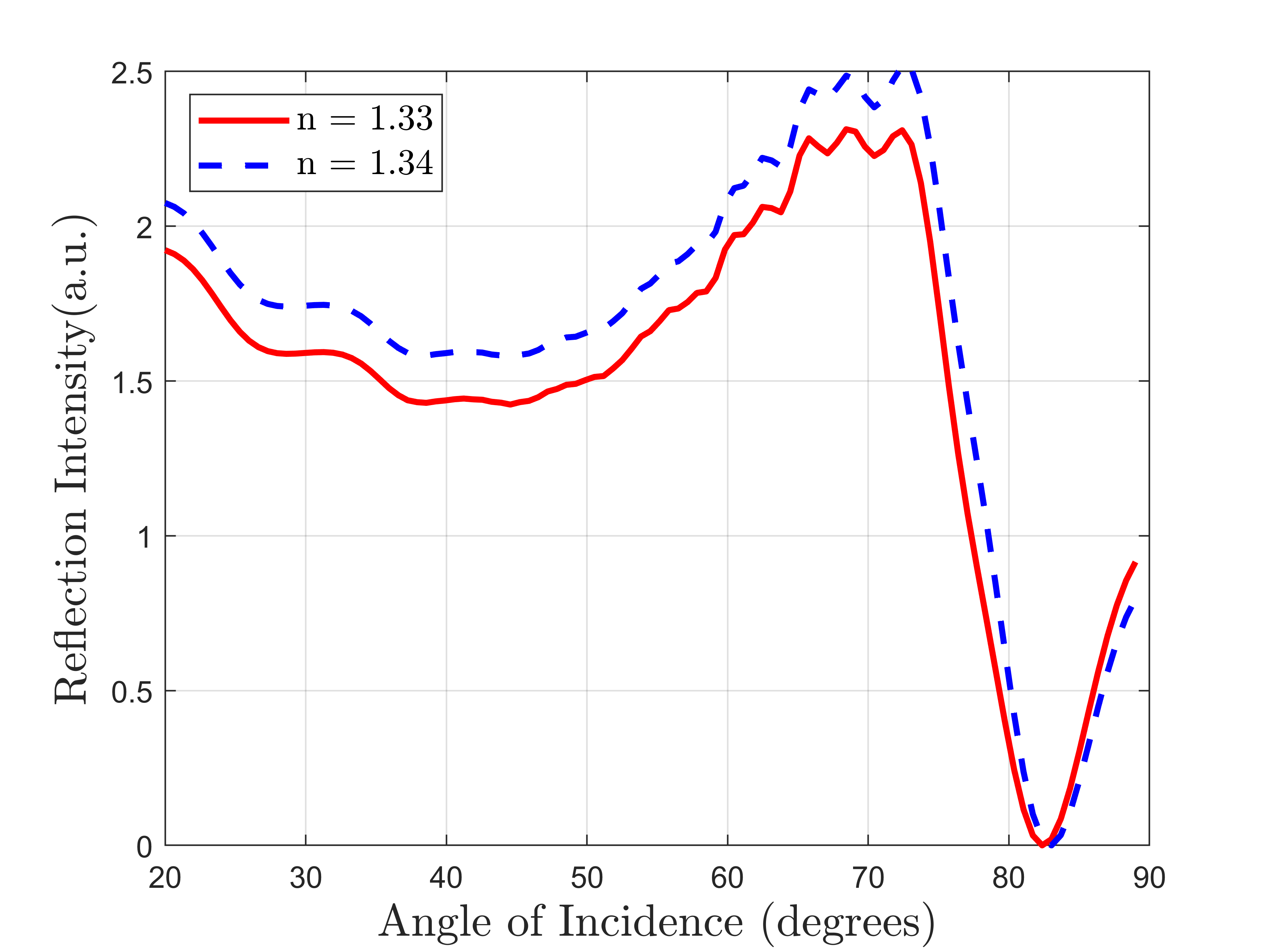}
\caption{Reflected light intensity versus source incident angle}
\label{FDTD}
\end{figure}

We also investigate our structure using an FDTD solver to verify our findings. As shown in Fig.~\ref{FDTD}, we get similar response for our proposed structure for the FDTD method as we get in transfer matrix method. This confirms robustness of our investigations.

\begin{figure}[h!]
\centering\includegraphics[width= 2 in]{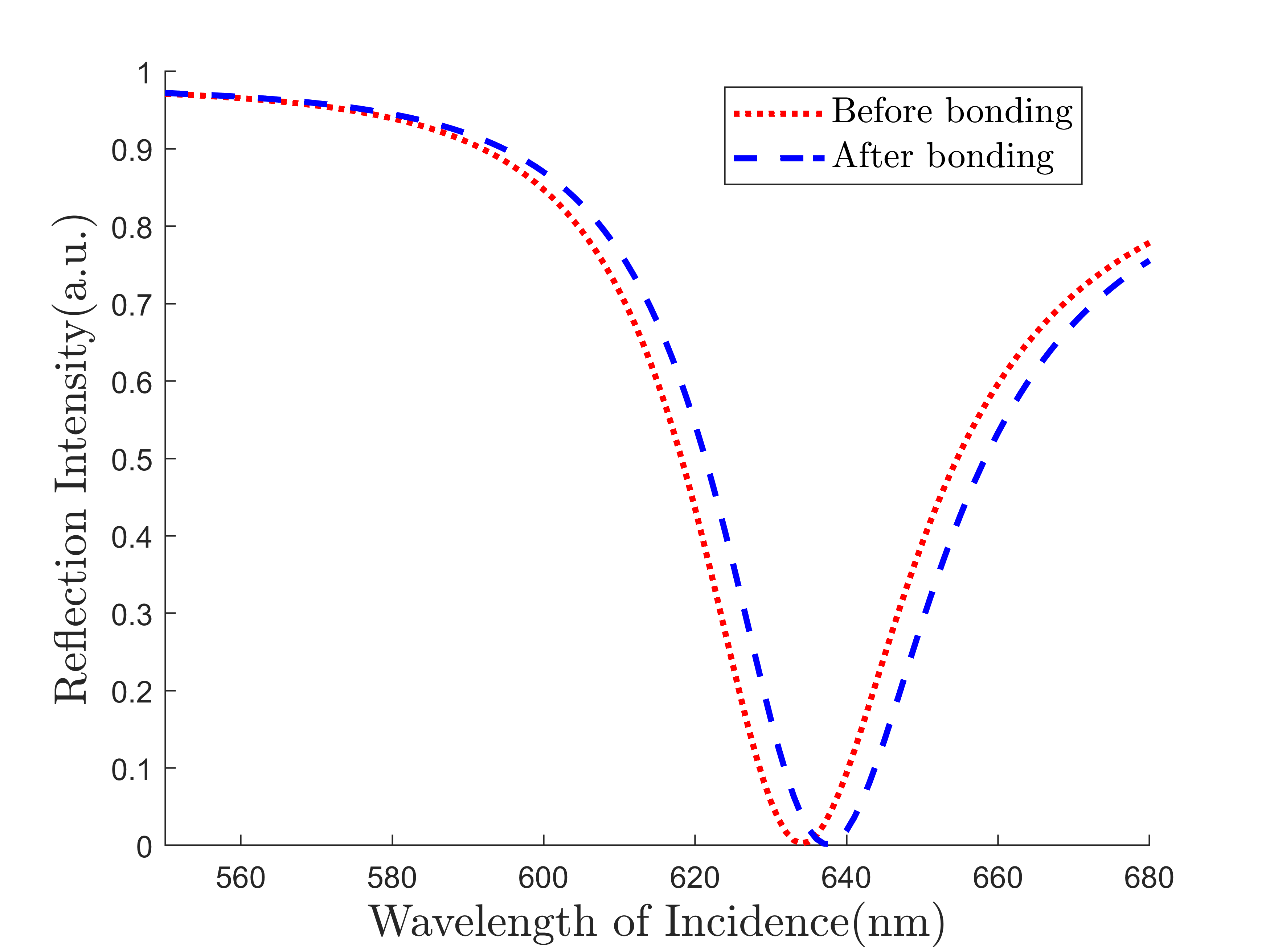}
\caption{Reflectance curve for wavelength interrogation of the proposed structure}
\label{web}
\end{figure}

Though our main goal is to maximize the performance in angular interrogation method, our structure shows promising results in wavelength interrogation process also. To investigate the proposed structure in wavelength interrogation method, we fixed the incident angle at 81 degree. Then we measure the reflected light at different wavelength and observe the reflectance dip. We vary the refractive index of the  sensing layer from 1.33 to 1.34 and observe the corresponding shift in resonance wavelength. We notice a significant red shift in the SPR curve as shown in Fig.~\ref{web}. We use the following equation to determine the sensitivity, 
\begin{equation}
    S_\lambda = \frac{d\lambda}{dn_s},
\end{equation}
where $d\lambda$ is the shift of resonance wavelength due to the change of refractive index of the sensing layer, $dn_s$. We get sensitivity as high as 300 $nm$/RIU.

\begin{figure}[h!]
\centering\includegraphics[width= 2 in]{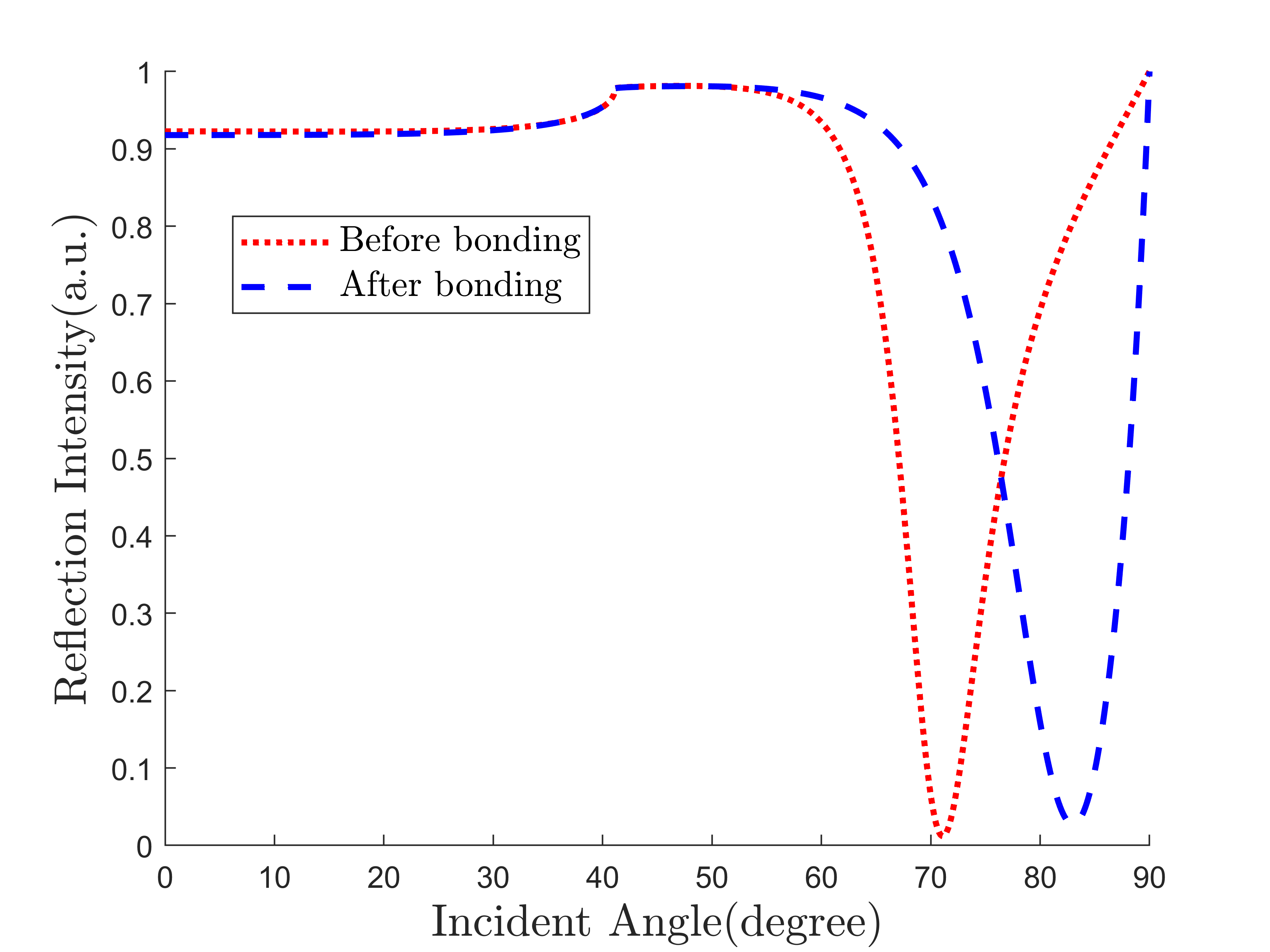}
\caption{Reflectance curve for detection of SARS-COV-2}
\label{cov}
\end{figure}

We now investigate the sensing performance of our structure for SARS-COV-2 detection. We use Thiol tethered DNA as a ligand layer as it has been proved to be a good receptor of SARS-COV-2. We collect data related to SARS-COV-2 from literature\cite{wang2020structural}. In the proposed sensing scheme, samples collected from human nasopharyngeal swabs are passed through the sensing channel in a liquid solution. When hybridization happens between SARS-CoV-2 RNA (RdRp-COVID sequence) from sample with thiol tethered DNA of receptor molecules, it causes  a significant SPR angle shift (as high as 12 degree) as demonstrated in  Fig.~\ref{cov}. Moreover, the reflectance dip is accompanied with very low Rmin (0.01587) and FWHM (11.86), which shows the prowess of our design as a potential plasmonic sensor for highly sensitive SARS-COV-2 detection. 
\subsection{Comparison with other reported structures}
Finally, we compare the performance parameters of our design with existing structures in literature. The performance metrics taken into account for comparison purposes are sensitivity, Rmin and FWHM. While sensitivity is the most important performance parameter under consideration, other two are also crucial for sensing accuracy. Hence, we define FOM according to Eqn.~7 as a benchmark for comparison. We show the findings in Table 2. From the table, we see that the proposed design shows higher sensitivity than majority of structures. While some designs have obtained better sensitivity compared to ours, they simultaneously suffer from higher Rmin and FWHM, and hence lower FOM. In terms of FOM, clearly our proposed structure outperforms other competitive counterparts. 
\begin{table}[h!]
\centering
\caption{Sensing performance comparison with other published results}
\fontsize{9}{9}\selectfont
\begin{tabular}{|c|c|c|c|c|}
\hline
\textbf{Sensor configuration} & \textbf{Sensitivity} & \textbf{FWHM} & \textbf{Rmin} & \textbf{FOM} \\ \hline
Proposed Structure     & 130.3                & 11.86         & \textbf{0.01587}       & \textbf{692.28}      \\ \hline
Au-Si-MX$_2$ \cite{st1}        & 147.88               & 16.2417       & 0.024099      & 377.81       \\ \hline
\begin{tabular}[c]{@{}c@{}}$TiO_2-SiO_2$-Ag-$MoS_2$-Graphene\cite{st2}\end{tabular}         & 98                   &  \textbf{1.102}         & Not reported  & Not reported \\ \hline
Au-$MoS_2$-Graphene\cite{st3}        & 89.29                & 6.80          & 0.025         & 525.2        \\ \hline
Cr-Ag-ITO\cite{st4}        & 68.77                & 1.64          & 0.0903        & 464.37       \\ \hline
ZnO-Ag-Au-Graphene\cite{st5}        & 76                   & 5             & 0.04          & 380          \\ \hline
ZnO-Au-$MoS_2$-Graphene\cite{st6}        & 101.58               & 6.73          & Not reported  & Not reported \\ \hline
Cu-$MoS_2$-Graphene\cite{st7}        & 79.12                & 2.89          & 0.0785        & 348.755      \\ \hline
Air-$MoS_2$-Al-$MoS_2$-Graphene\cite{st8}        & \textbf{190.83  }             & 19            & 0.04          & 251.09       \\ \hline
Mirrored bilayer of Au-$MoS_2$-Graphene\cite{st9}        & 75.2                 & 17            & 0.1           & 44.23        \\ \hline
Au-Graphene\cite{st10}       & 53.71                & 5             & 0.1252        & 85.79        \\ \hline
Chalcogenide Prism-Au-Graphene\cite{st11}       & 40                   & 2.5           & 0.1           & 160          \\ \hline
\end{tabular}%

\end{table}
\subsection{Robustness analysis}
For a sensor architecture to be successful in practice, ideally it should be able to provide consistent performance over a wide range of sensing molecule configurations. That is why, we also studied and analyzed our proposed framework for other ligand-ligate pairs. Notably, we observe similar improvement in performance for different receptor and sensing samples which demonstrates the robustness of our structure. We tabulate these results in Table 3.  
\begin{table}[h]
\centering
\fontsize{9}{9}\selectfont
\raisebox{0.95\height}{
\parbox{.35\linewidth}{
\caption{Robustness}

\begin{tabular}[h]{|c|c|}
\hline
\textbf{Ligand-Ligate Pair} & \textbf{Sensitivity} \\ \hline
BSA-Phospholipid            & 121.6                \\ \hline
BSA-Egg Yolk                & 121.46               \\ \hline
\end{tabular}
}}
\end{table}

\section{Conclusion}
We have analyzed an SPR based novel biosensor for label-free and real-time detection of SARS-COV-2. Extensive numerical simulations have been carried out for structural optimization and performance analysis. Transfer matrix method as well as FDTD technique have been employed to perform the simulations. Our results have shown promising indications for the usage of the proposed sensor configuration as a potential detector of SARS-COV-2. We have considered not only sensitivity but also other standard evaluation metrics such as FWHM, Rmin to demonstrate the performance of our design. Furthermore, we have defined a FOM to encapsulate all the performance parameters in a single evaluation quantity and compared with other existing structures. This investigation has demonstrated competitive effectiveness of our proposed structure over other designs. Moreover, the robustness of our sensor has been verified with two other different ligand-ligate pairs. The study is limited in the sense that the results are primarily numerical and experimental validation of such enhancements are required which is out of scope here and focus of future work. However, for experimental investigations, a well-designed and numerically evaluated sensor architecture is crucial, therein lies the significance of the presented study.

\bibliography{Reference}

\end{document}